\documentclass[10pt,conference]{IEEEtran}
\IEEEoverridecommandlockouts
\usepackage{cite}
\usepackage{amsmath,amssymb,amsfonts}
\usepackage{algorithmic}
\usepackage{graphicx}
\usepackage{textcomp}
\usepackage{xcolor}
\usepackage{subfigure}
\usepackage{collab}
\def\BibTeX{{\rm B\kern-.05em{\sc i\kern-.025em b}\kern-.08em
    T\kern-.1667em\lower.7ex\hbox{E}\kern-.125emX}}
    
\collabAuthor{zb}{teal}{Zhuangbin Chen}
\collabAuthor{yx}{purple}{Yuxin Su}
\begin{document}

\title{MTAD: Tools and Benchmarks for Multivariate Time Series Anomaly Detection
}

\author{\large Jinyang Liu\IEEEauthorrefmark{2}, Wenwei Gu\IEEEauthorrefmark{2}, Zhuangbin Chen\IEEEauthorrefmark{4}, Yichen Li\IEEEauthorrefmark{2}, Yuxin Su\IEEEauthorrefmark{4}, Michael R. Lyu\IEEEauthorrefmark{2}\\
\IEEEauthorblockA{
\IEEEauthorrefmark{2}The Chinese University of Hong Kong, Hong Kong SAR, China, \{jyliu, wwgu21, ycli21, lyu\}@cse.cuhk.edu.hk\\
\IEEEauthorrefmark{4}School of Software Engineering, Sun Yat-sen University, Zhuhai, China, \{chenzhb36, suyx35\}@mail.sysu.edu.cn\\
}
}

\maketitle

\begin{abstract}

Key Performance Indicators (KPIs) are essential time-series metrics for ensuring the reliability and stability of many software systems. They faithfully record runtime states to facilitate the understanding of anomalous system behaviors and provide informative clues for engineers to pinpoint the root causes. The unprecedented scale and complexity of modern software systems, however, make the volume of KPIs explode. Consequently, many traditional methods of KPI anomaly detection become impractical, which serves as a catalyst for the fast development of machine learning-based solutions in both academia and industry. However, there is currently a lack of rigorous comparison among these KPI anomaly detection methods, and re-implementation demands a non-trivial effort. Moreover, we observe that different works adopt independent evaluation processes with different metrics. Some of them may not fully reveal the capability of a model and some are creating an illusion of progress. To better understand the characteristics of different KPI anomaly detectors and address the evaluation issue, in this paper, we provide a comprehensive review and evaluation of twelve state-of-the-art methods, and propose a novel metric called \textit{salience}. Particularly, the selected methods include five traditional machine learning-based methods and seven deep learning-based methods. These methods are evaluated with five multivariate KPI datasets that are publicly available. A unified toolkit with easy-to-use interfaces is also released\footnote{https://github.com/OpsPAI/MTAD}. We report the benchmark results in terms of accuracy, salience, efficiency, and delay, which are of practical importance for industrial deployment. We believe our work can contribute as a basis for future academic research and industrial application.



\end{abstract}


\section{Introduction}
\label{sec: introduction}


In the past decades, modern software has been increasingly integrated into our daily lives, providing a variety of services such as instant messaging applications and search engines. With the growing number of customers worldwide, modern software systems have evolved into an unprecedented volume. Therefore, a short period of downtime could lead to significant revenue loss, especially for large-scale distributed systems~\cite{DBLP:conf/sigsoft/ChenKLZZXZYSXDG20,DBLP:conf/usenix/ZhangLXQ0QDYCCW19,DBLP:conf/icse/DangLH19,DBLP:journals/corr/abs-2009-07237,DBLP:conf/sigsoft/HeLLZLZ18, gu2023performance}.
To understand the health states of a system, Key Performance Indicators (KPIs) are recorded for continuously monitoring a system in different aspects, for example, CPU utilization, service response delay, network traffic, etc.

More often than not, a group of interdependent KPIs, i.e., multivariate KPIs, collectively can constitute a stronger indicator to capture the occurrence of system failures. Therefore, many studies have shifted to anomaly detection on multivariate KPIs. In recent years, there has been an explosion of interest in this field, and fruitful achievements have been made.

For example, Hundman et al.~\cite{DBLP:conf/kdd/HundmanCLCS18/lstm_ndt} applied long short-time memory network (LSTM) on telemetry data to detect the anomalies of spacecrafts in National Aeronautics and Space Administration (NASA). 
In addition, Borghesi et al.~\cite{borghesi2019anomaly} resorted to autoencoder to detect anomalies in high-performance computing systems.
Park et al.\cite{DBLP:journals/corr/abs-1711-00614/lstm-vae} combined LSTM with variational autoencoder (VAE) to detect anomalies on KPIs of robots.
Furthermore, Su et al. utilized a stochastic recurrent neural network to robustly model multivariate KPIs and achieve anomaly detection based on reconstruction probabilities. 
Li et al.~\cite{li2019mad} leveraged generative adversarial network (GAN) to concurrently consider the entire multivariate KPIs set to capture latent interactions among them. 
Zhang et al.~\cite{DBLP:conf/aaai/ZhangSCFLCNZCC19} proposed to model multivariate KPIs in a pair-wise manner to explicitly model the relation of the KPIs. 

However, as the evaluation process of different models and algorithms is conducted in an ad-hoc manner, the comparison among different methods may not reveal their capability. Moreover, the evaluation metrics, as well as the parameter tuning strategy, adopted also vary in different studies. These issues pose difficulties for researchers and practitioners in choosing appropriate methods that are suitable for the problems at hand.
Specifically, the general evaluation pipeline of an anomaly detection method goes as follows: given a segment of KPI observations, the model calculates an anomaly score which usually represents the probability of the KPI segment being anomalous. To make the definite decision, a threshold is often chosen. Anomalies are reported if the corresponding anomaly scores are greater than the threshold. In particular, common strategies for determining the threshold include Extreme Value Theory (EVT)~\cite{siffer2017anomaly} and iteration search within a feasible range. 

In the literature, two metrics are widely adopted to gauge a model's performance based on the prediction results: 1) Area Under the ROC Curve (AUC), 2) Precision, Recall, and F1 score. Particularly, based on the observation that human operators can know systems' misbehaviors if the model can alert an anomaly at least at one point of a contiguous anomaly segment, Xu et al.~\cite{DBLP:conf/www/XuCZLBLLZPFCWQ18/donut} proposed a way to adjust the prediction results, called point adjustment. As shown in Fig.~\ref{fig: point_adjustment}, point adjustment regards all the predictions within a consecutive range as true positive predictions if at least one anomaly in this range is reported by the anomaly detector. The adjusted results will be calculated to obtain an adjusted Precision, Recall, and F1 score, which are often better than their previous version. This modification essentially tells us the performance of a model in anomaly level instead of point-wise level. We argue that iteration search of threshold as well as point adjustment can significantly affect the performance evaluation. However, existing studies report the accuracy of computations based on different strategies. Hence, a benchmark for the state-of-the-art models with a unified evaluation protocol is desired.

Moreover, the existing evaluation process cannot be directly adopted because of the following drawbacks: 
(1) using the accuracy applied with point adjustment as the target for threshold iteration search tends to produce misleadingly higher anomaly detection evaluation results. 
(2) models may produce quite different anomaly scores even though they achieve similar accuracy.
(3) existing studies generally measure the performance in terms of accuracy. Nevertheless, an anomaly detection algorithm should be more comprehensively evaluated before deploying it in the production environment.
Moreover, though most of the authors kindly make the source code repositories public, reproducibility cannot be guaranteed because of the lack of detailed experimental configurations. 

To address these problems, we propose a unified benchmark protocol to comprehensively evaluate multivariate KPIs anomaly detection methods in terms of accuracy, \textit{salience}, delay, and efficiency. Especially within the protocol, a novel metric \textit{salience} is proposed to estimate how much a model can highlight the detected anomalies. Furthermore, we re-evaluate five general machine learning-based methods and seven state-of-the-art methods for multivariate KPIs anomaly detection on five widely-used public datasets.

Our experimental results reveal several thought-provoking phenomena: first, recent proposed deep learning-based methods do not necessarily outperform general machine learning-based methods in terms of accuracy. Second, deep learning-based methods are better at capturing long-lasting anomalies, while general machine learning-based methods tend to accurately report short-term anomalies. Third, we observe that methods with higher accuracy may not retain high salience, which indicates that a threshold should be carefully selected for the models. Fourth, most general machine learning methods can report anomalies earlier than deep learning-based methods. Fifth, except KNN and LOF, other general machine learning-based methods are efficient on both training and testing. AutoEncoder, LSTM, and LSTM\_VAE need more time to train but can achieve comparable efficiency on testing only with GPU acceleration.




%

To sum up, in this paper, we make the following major contributions:
\begin{itemize}
    
    \item We propose a comprehensive protocol with a novel metric \textit{salience}. Under this unified protocol, we extensively re-evaluate twelve methods, including five general machine learning-based methods and seven deep learning-based methods on five real-world multivariate KPIs datasets. 
    \item We implement and release an easy-to-use toolkit with a unified interface to automatically benchmark multivariate KPIs anomaly detection methods. We believe this toolkit can benefit both academia and industry for developing advanced anomaly detection algorithms. 
\end{itemize}

\section{Background}
\label{sec: background}

\subsection{Overview}
To ensure the reliability of large-scale systems, KPIs are usually utilized to monitor different components of the system. In doing this, the run-time status can be recorded by the KPIs based on a particular sampling rate, e.g., one minute per point, where each sample is called an observation for this moment. Generally, there are univariate KPIs and multivariate KPIs. Univariate KPI is a single KPI associated with a particular property, e.g., \textit{cpu 
temperature} of a physical machine. In addition, multivariate KPIs are a group of univariate KPIs utilized to monitor an entity, e.g., a virtual machine (VM). Sophisticated and sparse dependencies may exist within the multivariate KPIs because different components of an entity are tightly coupled and interact with each other. For example, if the visit for a cluster gateway node increases, the node will present increasing \textit{cpu utilization} because of more computation. Moreover, the network KPIs such as \textit{number of packets in} and \textit{memory utilization} can simultaneously increase. 

Anomaly detection is to detect unexpected abnormal patterns which differ from normal cases. Such patterns existing in multivariate KPIs are more complicated than univariate KPIs because the KPIs are correlated with each other based on their physical dependency\cite{DBLP:conf/icdm/ZhaoWDHCTXBTZ20}\cite{deng2021graph}.
Therefore, how to capture such dependency of the multivariate KPIs effectively is crucial for an anomaly detection method and remains a challenging research problem. To address this problem, many innovative methods are proposed recently. In this paper, we focus on benchmark the state-of-the-art anomaly detection methods on multivariate KPIs. In the following part of this section, we formally define multivariate KPIs anomaly detection and introduce representative methods.

\subsection{Problem Statement}
Multivariate KPIs are a group of correlated univariate KPIs used to monitor different aspects of an entity. 
The input of an anomaly detector is the multivariate KPIs, represented as a matrix $X\in \mathbb{R}^{n\times m}$. In this matrix, the $i^{th}$ row is the $i^{th}$ observation for all the KPIs, i.e., $X_i \in\mathbb{R}^{n}$. In addition, the $j^{th}$ column is a particular KPI of the entity, e.g., $X^j\in\mathbb{R}^{n}$ is the $j^{th}$ KPI containing all the observations.  The vector $y\in \mathbb{R}$ is used to denote the anomaly label for all the observations. Specifically, $y_i=1$ if the entity is anomalous at the $i^{th}$ observation, otherwise, $y_i=0$. In particular, we consider the anomaly at the entity level, namely, we consider all KPIs at the same time to detect the entity anomaly, rather than a specific KPI.
The output of an anomaly detection method is an anomaly score vector $s\in \mathbb{R}$, in which a higher score indicates a higher possibility for an anomaly. An effective anomaly detector should achieve higher anomaly scores for those anomalous observations and lower anomaly scores for the normal ones. After obtaining the anomaly scores, a threshold $\theta$ is selected to determine if an observation is anomalous or not. Generally, the $i^{th}$ observation is reported as anomalous if $s_{i}\geq \theta$, and vice versa, which yields a prediction vector $\hat{y}$ associated with anomaly score $s$.

\subsection{Representative Methods}
In this section, we summarize the representative methods that are evaluated in this work. In total, we evaluate five general machine learning-based methods and seven deep learning-based methods.

\subsubsection{Traditional Machine Learning-based Methods}

The topic of anomaly detection has been widely studied\cite{DBLP:journals/corr/abs-2004-00433/univariate_survey}. Many typical anomaly detectors have been developed based on machine learning technologies, which can be utilized to find general anomalies for various data types. We describe the selected models as follows:

\begin{itemize}
    \item \textbf{KNN} K-nearest neighbor (KNN) is a distance-based anomaly detection method~\cite{ramaswamy2000efficient}. In the multivariate KPIs, for each observation, the distance of its $k^{th}$ nearest neighbor is computed. After that, all the observations are ranked based on the distances. Finally, those observations with larger distances are more likely to be anomalies. Intuitively, the data point (i.e., observation) that is distant from most of the other points is an anomaly.
    \item \textbf{LOF} The density-based anomaly detection method LOF assigns a local outlier factor (LOF) for each input observation to detect anomalies~\cite{breunig2000lof}. Similar to KNN, the LOF of a data point is computed based on the distance to its K-nearest neighbor as well as the distances of its neighbors. By considering the relative density of the point itself and the neighbors, anomaly scores could be more effectively computed.

    \item \textbf{PCA} Principal component analysis (PCA) is a widely-adopted dimension reduction method, where a covariance matrix is computed. The covariance matrix can be decomposed into eigenvectors with eigenvalues. Those eigenvectors associated with larger eigenvalues can capture the majority of the data, i.e., the normal part. Therefore, the anomalous points are more obvious on the hyperplane formed by the eigenvectors with smaller eigenvalues. Therefore, anomaly scores can be obtained as the sum of the projected distance on all eigenvectors.

    \item \textbf{iForest} Isolation forest (iForest) is an ensemble-based anomaly detection method~\cite{DBLP:conf/icdm/LiuTZ08/isolation_forest}, which explicitly isolates anomalies based on subsets of input features by isolation trees. For multivariate KPIs, first, a subset of all KPIs is randomly selected. Then, different observations are partitioned according to the value range of the KPIs, which generate a tree. After convergence, the observations are isolated to leaf nodes of the tree. Isolation forest could be produced by repeating the procedure multiple times. If an observation is anomalous, it should be easier to be isolated. Hence, the height of a leaf node could be used to measure the anomaly score of an observation.

    \item \textbf{LODA} LODA is a lightweight online anomaly detector based on an ensemble of very weak detectors~\cite{pevny2016loda}. LODA consists of a collection of one-dimensional histograms, each of which projects the input data onto a single projection vector to approximate the probability density. Anomaly scores could be computed based on the probability density. Hence, if an observation has a low probability density, a high anomaly score can be computed.
    
\end{itemize}

\subsubsection{Deep Learning-based Methods}
Moreover, recently, with the successful application in computer vision, deep learning has been introduced to multivariate KPIs anomaly detection~\cite{DBLP:conf/kdd/SuZNLSP19/omni}\cite{deng2021graph}. We select state-of-the-art models for evaluation and describe them as follows:

\begin{itemize}
    \item \textbf{AE} Autoencoder (AE) is an encoder-decoder neural network architecture. The encoder compresses the input data into a hidden vector in a low-dimensional space, from which the decoder attempts to reconstruct the original input data. The training is conducted by minimizing the reconstruction error. When performing anomaly detection, the reconstruction error is used to measure the anomaly scores~\cite{borghesi2019anomaly}. Intuitively, anomalous observations are rarely seen in the training process. Hence, a larger reconstruction error indicates a higher anomaly score.
    \item \textbf{LSTM} Long short-term memory (LSTM) network adopts a recurrent structure to process sequential data. During the training process, preceding observations are fed into LSTM to predict the next observation, where the prediction error is minimized. To detect anomalies, the next actual prediction is compared with the predicted one. If the discrepancy is large, the input observations may be anomalous, which follows a similar philosophy as AE. This method has been used to detect anomalies within spacecraft by NASA~\cite{DBLP:conf/kdd/HundmanCLCS18/lstm_ndt}.
    \item \textbf{LSTM-VAE} The method LSTM-based variational autoencoder (LSTM-VAE) combines LSTM and variational AE to detect anomalies~\cite{DBLP:journals/corr/abs-1711-00614/lstm-vae}. In doing this, the underlying distribution of multivariate KPIs could be well modeled. Then, the anomaly scores could be estimated by the negative log-likelyhood with respect to the distribution.
    \item \textbf{DAGMM} Deep Autoencoding Gaussian Mixture Model (DAGMM) is also based on AE~\cite{DBLP:conf/iclr/ZongSMCLCC18/dagmm}. DAGMM attempts to jointly optimize the parameters of an encoder and an estimation network. With this method, DAGMM can generate a more effective internal representation for anomaly detection. Then, the anomaly score could be estimated by the likelihood of an input observation.
    \item \textbf{MAD-GAN} MAD-GAN is a multivariate anomaly detection method based on generative adversarial networks (GAN)~\cite{li2019mad}. The base model of MAD-GAN is an LSTM network, which captures the temporal correlation of KPIs. Besides, MAD-GAN proposes to use the DR-score based on discrimination and reconstruction to estimate anomaly scores.
    \item \textbf{MSCRED} MSCRED explores the inter-correlation explicitly between different pairs of multivariate KPIs to capture the sophisticated dependencies~\cite{DBLP:conf/aaai/ZhangSCFLCNZCC19}. Specifically, a signature matrix is first built by combining all the KPIs in a pair-wise manner. Then, a convolutional LSTM (ConvLSTM) is developed to capture the temporal patterns. Finally, a convolutional decoder is utilized to measure the anomaly scores based on reconstruction errors.
    \item \textbf{OmniAnomaly} OmniAnomaly is a stochastic recurrent neural network that can model multivariate KPIs robustly~\cite{DBLP:conf/kdd/SuZNLSP19/omni}. OmniAnomaly first learns normal patterns of multivariate KPIs by using the stochastic variable connection and planar normalizing flow. After reconstruction, anomaly scores are estimated by reconstructing probabilities.
\end{itemize}

\subsection{Representative Evaluation Protocol}
In the literature, different methods are evaluated under different evaluation protocols. The protocols are general in a two-stage manner: threshold selection and accuracy computation. Specifically, in the first stage, after obtaining anomaly score $s$ from an anomaly detector, a threshold should be selected based on domain knowledge, the distribution of anomaly scores or searching. With this threshold, anomaly scores are transformed to anomaly prediction $\hat{y}$.
In the second stage, the anomaly prediction is compared with the ground truth $y$ to compute accuracy, which is used to estimate the performance of an anomaly detector.
The two stages are elaborated in detail as follows.

\subsubsection{Threshold Selection} 
Generally, there are three categories of methods to select a proper threshold: (1) manual selection based on domain knowledge~\cite{DBLP:journals/corr/abs-1711-00614/lstm-vae}\cite{DBLP:conf/aaai/ZhangSCFLCNZCC19}\cite{ren2019time}. This requires labor-intensive inspection on the anomaly scores and sufficient domain knowledge. Hence it is not scalable if a large number of KPIs are monitored in a large-scale system.
(2) automatic selection based on the distribution of anomaly scores. To be more specific, the principle of Extreme Value Theory (EVT)~\cite{siffer2017anomaly} is generally adopted\cite{DBLP:conf/kdd/HundmanCLCS18/lstm_ndt}\cite{DBLP:conf/icdm/ZhaoWDHCTXBTZ20}\cite{DBLP:conf/kdd/SuZNLSP19/omni}. The object of EVT is to find the law of extreme values without making any assumption on data distribution. The only required parameter is a risk ratio to determine the number of false positive anomalies. EVT can set a threshold without any ground truth labels or experienced experts. However, it still requires the risk ratio to be manually set and may suggest a suboptimal threshold.
(3) searching the optimal threshold within a range. This is also a widely used method to estimate the best performance of a model~\cite{li2019mad}\cite{DBLP:conf/kdd/AudibertMGMZ20/usad}. Specifically, the thresholds are enumerated within a range with the aim to achieve the best accuracy in the labeled test data. Then, the best accuracy associated with the threshold is reported as the performance of the evaluated model.

\subsubsection{Accuracy Computation}
The accuracy of a method is usually evaluated in terms of area under the curve(AUC), precision (PC), recall (RC), and F1 score. In particular, AUC is measured by the probability (ranging from 0 to 1) that a randomly chosen anomalous sample is ranked higher than the randomly chosen normal samples, which could be computed directly from anomaly scores and ground truth labels. PC, RC and F1 score are computed according to Equation~\ref{Equ: pc_rc_f1}. In this equation, TP (true positive) is the number of true anomalies that are reported by the model as anomalies. FP (false positive) is the number of normal samples that are indicated as anomalies by the model. FN (false negative) is the number of anomalous samples that are predicted as normal ones. Therefore, PC estimates the ability of a model to confidently predict the anomalies and RC measures how many anomalies are missed by the model. F1 score is a harmonic mean of PC and RC. Particularly, all the deep learning-based models we selected in this benchmark used the F1 score in their original papers.

$$
    PC = \frac{TP}{TP + FP}\quad
    RC = \frac{TP}{TP + FN}\quad
    F1 = \frac{2 \cdot PC \cdot RC}{PC + RC}\label{Equ: pc_rc_f1}
$$

\begin{figure}[h]
    \centering
    \includegraphics[width=1\columnwidth]{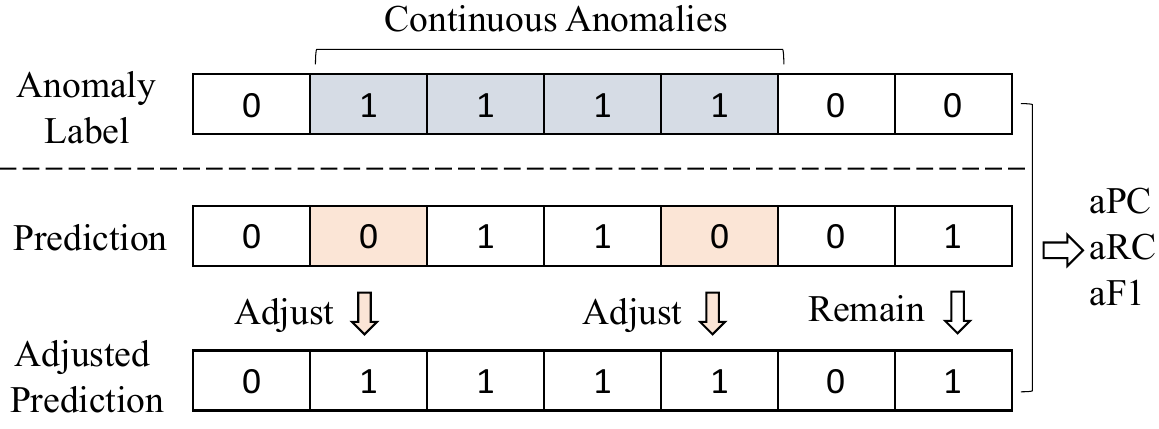}
    \caption{An Example of Point Adjustment}
    \label{fig: point_adjustment}
    \end{figure}

Considering that anomalies may last in real-world applications, \textit{point adjustment} is proposed to facilitate the accuracy computation~\cite{DBLP:conf/www/XuCZLBLLZPFCWQ18/donut}. To be more specific, if an algorithm reports an anomaly within a continuous anomaly segment, the predictions within the segment are adjusted to anomalies though the original prediction may be normal. Besides, the points outside the anomalous segment remain unchanged. Figure~\ref{fig: point_adjustment} shows an example of point adjustment. The predictions are compared with the anomaly labels. In the range of the continuous anomalies, the zeros in the prediction (i.e., normal points) are adjusted to ones (i.e., anomalies). However, the last one in the prediction remains unchanged, which is a false positive. Finally, the adjusted F1 (aF1), adjusted PC (aPC), and adjusted RC (aRC) are computed based on the adjusted predictions, respectively. 

In recent studies, the accuracy with point adjustment is usually adopted to evaluate the performance of anomaly detection methods on KPIs~\cite{DBLP:conf/www/XuCZLBLLZPFCWQ18/donut}\cite{DBLP:conf/icdm/ZhaoWDHCTXBTZ20}\cite{DBLP:conf/kdd/SuZNLSP19/omni}\cite{DBLP:conf/kdd/AudibertMGMZ20/usad}\cite{DBLP:conf/kdd/RenXWYHKXYTZ19/microsoft}. In this paper, we argue that point adjustment may potentially lead to higher F1 score that is against to the real performance of the model, which will be elaborated in the following section.

\section{Motivation}
\label{sec: motivation}

Multivariate KPIs anomaly detection methods are generally evaluated under the previous evaluation protocol. However, we argue that the models with high accuracy under this protocol do not necessarily perform well in the real-world application for the following issues:

\textbf{Issue 1} Point adjustment potentially dominates false positive observations, which may yield misleading higher F1 scores. Specifically, as for a number of observations, if a model reports an anomaly within a continuous anomalous range, all other points in this range are regarded as true positive predictions, regardless of the points that the model misses in the range. In doing this, the model may be over-rewarded for true positive predictions and false positive ones are dominant because they remain unchanged after point adjustment (the last point in the prediction vector in Fig.~\ref{fig: point_adjustment}). Intuitively, a true positive prediction in an anomalous range may incur a number of false negative points to be adjusted. In contrast, false positive predictions are not adjusted, which becomes more tolerable as a consequence.

\begin{figure}[h]
    \centering
    \includegraphics[width=0.95\columnwidth]{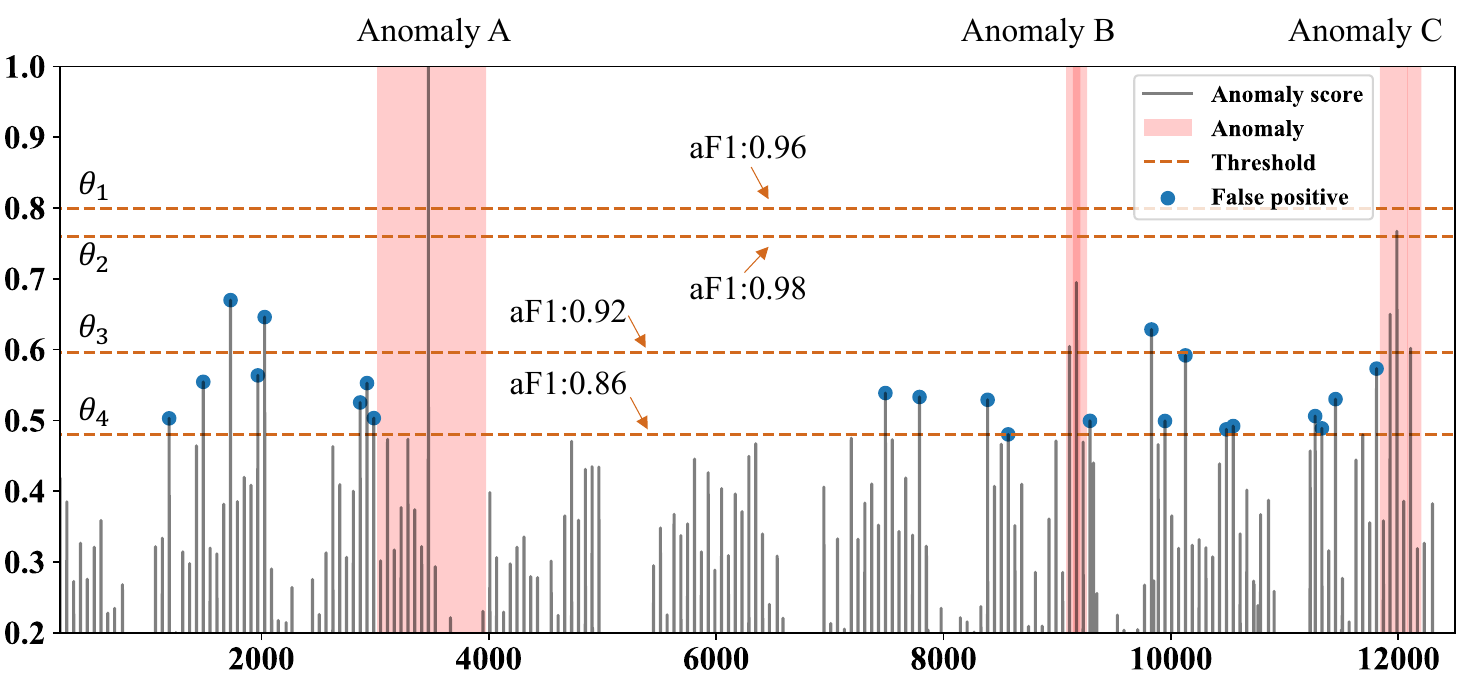}
    \caption{Prediction Results of LSTM on SMD Dataset (Machine-3-2)}
    \label{fig: af1_concrete_example}
    \end{figure}

\begin{figure}[h]
    \centering
    \includegraphics[scale=0.36]{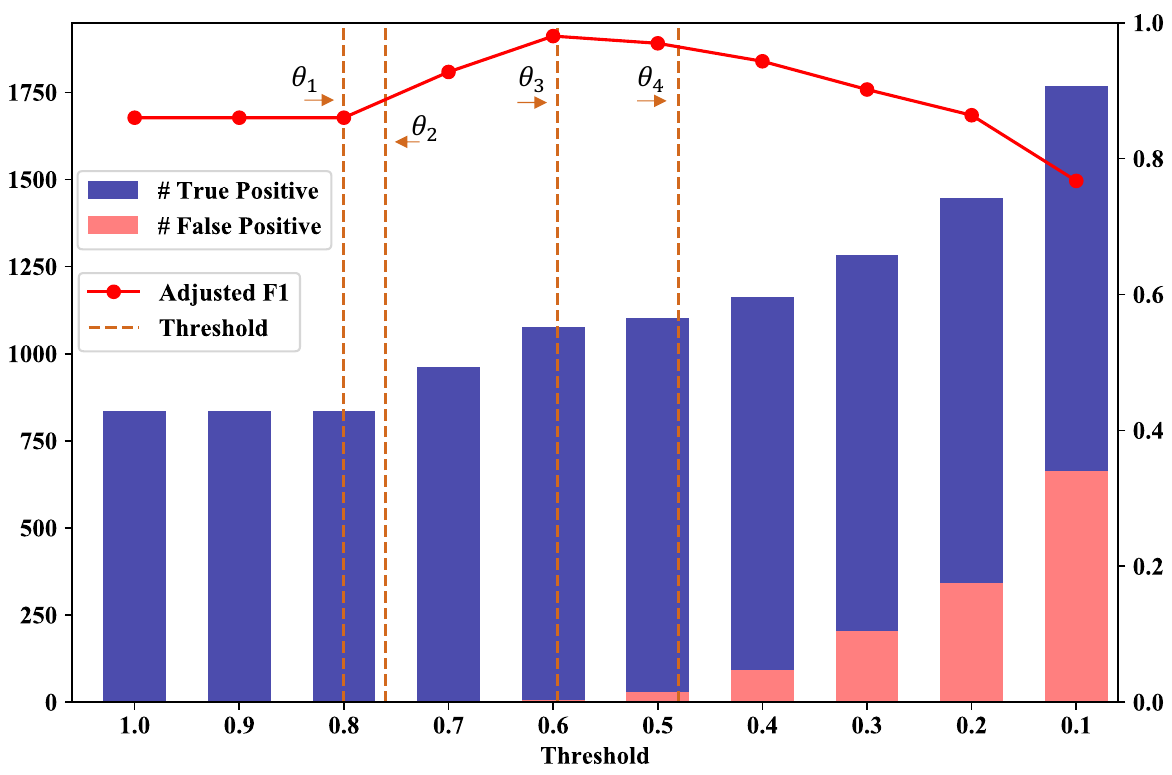}
    \caption{Number of True/False Positive Predictions v.s. Different Threshold: Because the true positive predictions in Anomaly A dominate the false positive ones, the adjusted F1 scores do not decrease significantly according to different thresholds.}
    \label{fig: iter_adjust_f1_ratio}
    \end{figure}

Fig.~\ref{fig: af1_concrete_example} shows a real-world example for this issue. We obtain the anomaly scores after running the LSTM anomaly detector on the machine-3-2 of the SMD dataset collected from industrial scenario~\cite{DBLP:conf/kdd/SuZNLSP19/omni}. We mark three anomalous areas with the light red shadows (Anomaly A to C) and aF1 scores associated with four thresholds from high to low, namely, $\theta_1$ to $\theta_4$. Anomaly A can be easily captured by $\theta_1$, which achieves a high aF1 0.96. Moving from $\theta_1$ to $\theta_2$ captures Anomaly C, which increases aF1 to 0.98. Moreover, moving to $\theta_3$ and $\theta_4$ incurs more false positive predictions, which decreases the aF1. However, though many false positive predictions are made, the aF1 does not decrease significantly and is as high as 0.86. Fig.~\ref{fig: iter_adjust_f1_ratio} shows concrete numbers of true and false positive predictions for different thresholds. With the threshold decreasing from 1 to 0.1, though false positive cases are increasing, true positive predictions are dominant, therefore, the aF1 curve remains at high values. To summarize, in the case that long-lasting continuous anomalies exist, utilizing threshold searching with the object to maximize aF1 can potentially lead to a high aF1 score, where false positive cases are dominated.

\begin{figure}[h]
    \centering
    \includegraphics[scale=0.35]{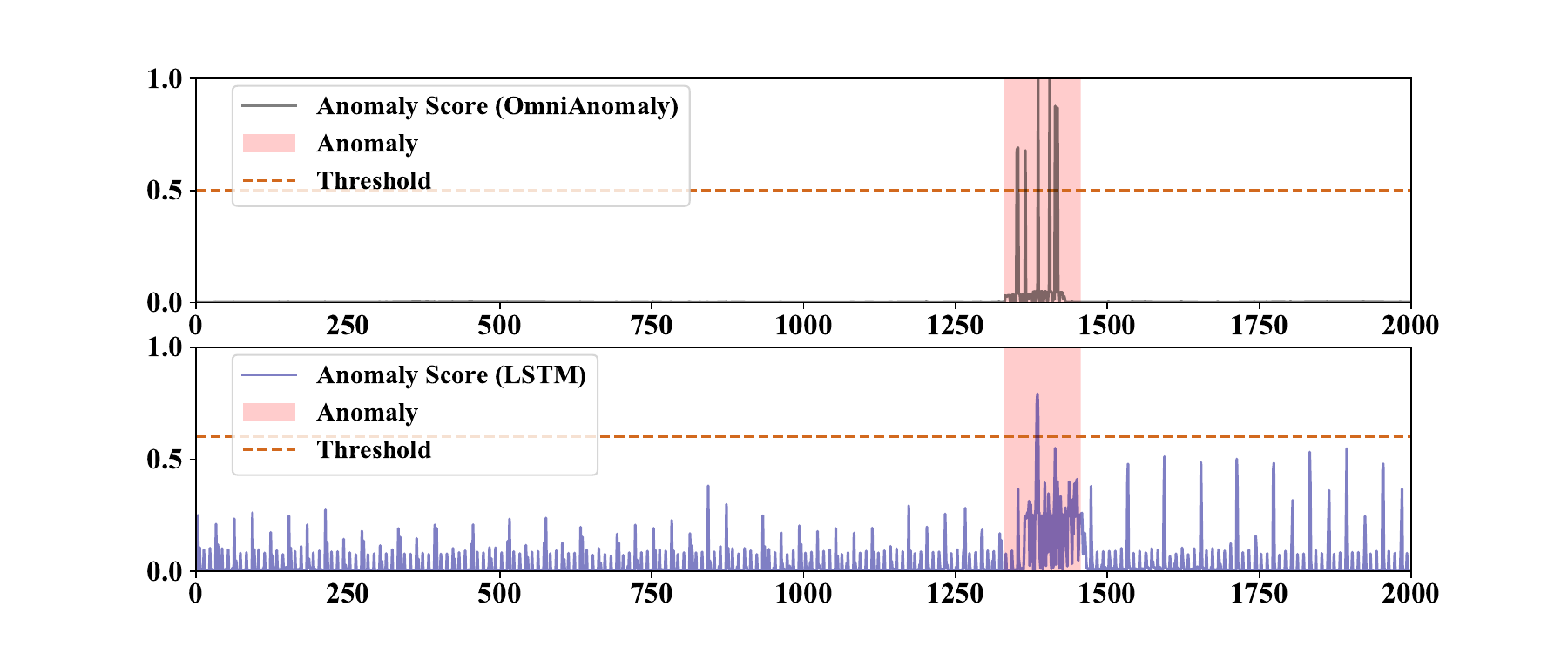}
    \caption{Prediction Results of OmniAnomaly and LSTM on SMD Dataset (Machine-3-9): Both of the methods successfully predict higher anomaly scores within the anomalous area. However, OmniAnomaly is more practical because the anomaly scores are more obvious.}
    \label{fig: two_same_f1}
    \end{figure}

\textbf{Issue 2} Models with similar accuracy (e.g., F1 or aF1 scores) may behave differently in real-world adoption. Fig.~\ref{fig: two_same_f1} shows the anomaly scores obtained after running OmniAnomaly and LSTM on the SMD dataset (machine-3-9). In this case, both of the two algorithms can successfully report the anomalous range with corresponding selected thresholds. Therefore, their accuracy in terms of aF1 is the same. However, OmniAnomaly is more practical and easier to use because the anomaly scores in the anomalous range are more salient than the outside range. In contrast, LSTM presents only slightly higher anomaly scores in the anomalous range than the outside range. Namely, in this case, OmniAnomaly gives more choices for threshold selection to practitioners than LSTM, because one should carefully choose thresholds for LSTM to avoid incurring more false positive predictions. However, the current evaluation protocol based on accuracy cannot describe such characteristics. An additional metric showing how many anomalous scores stand out compared to the normal values is desired. Therefore, we propose a novel metric salience, which will be elaborated in Section~\ref{sec: framework}.

\textbf{Issue 3} The evaluation of current studies is not comprehensive and fails to consider crucial aspects for industrial deployment. Besides accuracy, the metric \textit{delay} and \textit{efficiency} should also be included in the evaluation protocol. First, delay measures how timely an anomaly could be detected, which is important because a short period of downtime for a large-scale system could cause significant revenue loss~\cite{DBLP:conf/sigsoft/ChenKLZZXZYSXDG20}. The practitioners could have more time to locate the root cause if an anomaly could be reported earlier. Second, the volume of KPIs is generally large with the scale of software growing. Hence, the efficiency of an algorithm should be measured to provide guidance for industrial use. We formally define delay and efficiency in our following benchmark protocol.

To summarize, the current evaluation protocol based on accuracy is not comprehensive and potentially presents higher F1 scores, which makes it empirical to select a well-performed anomaly detection algorithm for deployment. To fill in this gap, we propose to comprehensively evaluate multivariate KPIs anomaly detection algorithms in the following benchmark protocol.

\section{Our benchmark Protocol}
\label{sec: framework}

In this section, we introduce our benchmark protocol. Our goal is to provide a comprehensive evaluation framework, which addresses the aforementioned issues and advises the real-world application of anomaly detection algorithms over multivariate KPIs.

\subsection{Overview}

In our benchmark protocol, we evaluate anomaly detectors in terms of accuracy, \textit{salience}, delay and efficiency. In particular, we propose the novel metric salience to measure how obvious the anomalous area (i.e., anomaly scores within the range of anomaly happening) is compared with the normal area. The aim of using salience is to make a supplement to accuracy to help practitioners select the most suitable in a specific scenario. We elaborate on each metric in detail as follows.

\subsection{Accuracy}
The accuracy of an anomaly detection algorithm is generally estimated by F1 scores~\cite{DBLP:conf/issre/HeZHL16/loglizer}\cite{DBLP:conf/ccs/Du0ZS17/deeplog}. However, for multivariate KPIs anomaly detection, F1 scores could be computed in different ways by combining different threshold selection strategies (i.e., searching or automatic selection with EVT~\cite{DBLP:conf/kdd/SifferFTL17/EVT}) and whether to use point adjustment. To be more specific, we first normalize the anomaly score $s$ to $\hat{s} \in [0,1]$ with min-max normalization shown in Equation~\ref{equ: normalization}. 
\begin{equation}
    \hat{s} = \frac{s - \min(s)}{\max(s) - \min(s)}
    \label{equ: normalization}
\end{equation}
Then, if we utilize EVT to compute a threshold $\theta_e$, the prediction vector $\hat{y}_e = \{I(\hat{s}_i >= \theta_e)| i \in [1,2,...,n]\}$ could be obtained, where $I(\cdot)$ is the indicator function that outputs $1$ if the input expression is true and $0$ otherwise. In particular, for the algorithm that regards smaller anomaly scores as anomalies, we take $\hat{s} = 1 - \hat{s}$ to make it consistent with our computation. Furthermore, we can apply point adjustment on $\hat{y}_e$ to obtain $\hat{y}^*_e$ according to anomaly label vector $y$. Hence, we denote the F1 scores based on $\hat{y}_e$ and $\hat{y}^*_e$ as $F_1$ and $F_1^*$, respectively. Besides, we can search the optimal threshold by enumerating the threshold from 0 to 1 with the step 0.01. For each enumeration, a prediction vector is generated to compute an F1 score (without adjustment). We finally keep the threshold $\theta_s$ and prediction vector $\hat{y}_s$ associated with the highest F1 score. Likewise, we can adopt point adjustment on $\hat{y}_s$ to generate $\hat{y}^*_s$, where corresponding F1 scores $\hat{F}_1$ and $\hat{F}_1^*$ can be obtained. In this procedure, we prevent using adjusted F1 score as the optimizing object while searching the threshold, which may lead to misleading higher F1 scores. Instead, we use the threshold $\theta_s$ that achieves the best F1 score in the searching process and generate the adjusted predictions based on $\theta_s$.


\subsection{Salience}
Algorithms with similar accuracy could present different characteristics within their anomaly scores, as shown in Fig.~\ref{fig: two_same_f1}. Intuitively, both of them could report the anomalies, but the significance is different. In this case, the algorithm with higher significance is more practical. Therefore, we propose the metric \textit{salience} to describe this characteristic quantitatively. Specifically, Salience computation is based on the normalized anomaly scores $\hat{s}$ and label $y$, which consists of two steps: \textit{support points computation} and \textit{discrepancy measurement}.

\begin{figure}[h]
    \centering
    \includegraphics[scale=0.36]{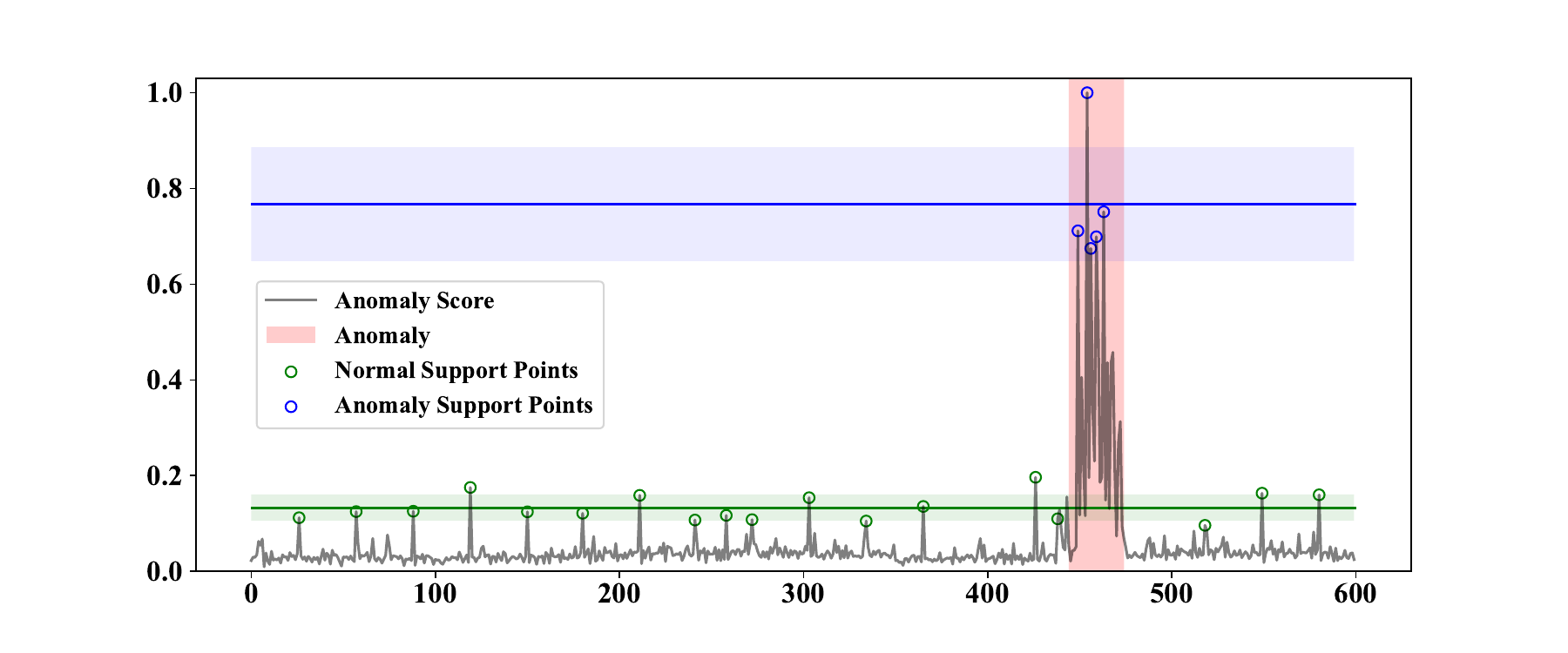}
    \caption{An Example of Salience Computation}
    \label{fig: salience_demo_case}
    \end{figure}

\subsubsection{Support points computation} 
To measure the significance of anomaly scores within the anomalous areas (marked with red shadow in Fig.~\ref{fig: salience_demo_case}) compared with other areas (the unshaded areas), we need to compare the discrepancy of the significant parts (e.g., spikes) of the two areas regardless of the lower points, because we generally report the points with corresponding anomaly scores higher than a threshold.

To achieve this, we first conduct support points computation. Intuitively, support points are those points that are more significant than others within a specific area (i.e., the anomalous or normal area). As shown in Fig.~\ref{fig: salience_demo_case}, we use anomaly support points (ASP) and normal support points (NSP) to denote the support points in the anomalous and normal range, respectively. To compute ASP and NSP, we first divide the anomaly scores $\hat{s}$ to an anomalous set $\hat{s}_a = \{\hat{s}_i | i\in \{y_i=1\}\}$ and a normal set $\hat{s}_n= \{\hat{s}_i | i\in \{y_i=0\}\}$. Then, we perform the following two-class agglomerative clustering on the two sets separately.

Given an anomalous set $\hat{s}_a$, each element is initialized as a single cluster itself. Then, we merge the closest two clusters. To compute the closest clusters, we adopt complete linkage~\cite{wiki_complete_linkage} to measure the distance between two clusters, which is defined as the maximum distance between any two elements of the clusters. We stop merging when only two clusters are left.
After that, we regard the points in the cluster with higher average score as ASP, where the average score is calculated by the summation of the scores in the cluster divide by the number of points. Likewise, we conduct the clustering process on the normal set $\hat{s}_n$ to obtain NSP.

\subsubsection{Discrepancy measurement}
We then compute the statistical discrepancy between ASP and NSP as the salience. Particularly, we first compute the mean values for ASP and NSP as $\mu_a$ and $\mu_n$, respectively. Then, we measure salience by computing the weighted discrepancy of $\mu_a$ and $\mu_n$ with the following Equation~\ref{equ: salience}:
\begin{eqnarray}
    \text{salience} &=& w_a \times \mu_a - w_n \times \mu_n \label{equ: salience} \\
    w_a &=& \phi(\frac{|A|}{|A+|N|}) \\ 
    w_n &=& \phi(\frac{|N|}{|A+|N|}), 
\end{eqnarray}
where $|A|$ and $|N|$ denotes the number of points contained in ASP and NSP, respectively. $\phi(x) = \frac{1}{1+e^{-x}}$ is the sigmoid function used for smoothing purpose. Hence, $w_a$ and $w_n$ depict the weights of the mean values, i.e., $\mu_a$ and $\mu_n$. Intuitively, if a mean value is computed based on more supporting points, it has a larger contribution in computing salience.

In our benchmark, salience works as a supplement for accuracy to automatically distinguish the more superior anomaly detector when two detectors have similar accuracies. In real-world practice, we believe practitioners can benefit from salience to understand, maintain and improve an anomaly detection system.

\subsection{Delay}
Delay is an essential metric to estimate how timely a model could detect anomalies. An algorithm with lower delay is more practical to industrial applications, which allows a timely fix in order to prevent potential system fault. For each continuous anomalous segment, the delay could be computed as the difference between the start of the anomaly and the first predicted anomalous point. Delay is also affected by the sampling rate of KPIs, however, in this paper, we report delay in terms of the number of lagged points instead of concrete time, which can be easily generalized to different monitoring systems. Moreover, for multivariate KPIs with multiple anomalous segments, we report the summation of delays of all segments.

\subsection{Efficiency}
Most of the recent machine learning models~\cite{zhao2019pyod} consist of the training phase and testing phase. Hence, the efficiency of an algorithm should be measured based on both phases, namely, the time required to train to convergence and the time required to conduct testing. Higher training efficiency allows the model to be frequently updated to catch the latest pattern of KPIs to alleviate concept drift~\cite{DBLP:journals/tkde/LuLDGGZ19}. Besides, higher testing efficiency indicates that a model can process online streaming data with higher throughput.

\subsection{Tool Implementation and Reproducibility}
Current multivariate KPIs anomaly detection researchers kindly release their source code repositories, but they are organized with different input/output interfaces based on datasets with different formats, which makes it labor-intensive for customized applications. For ease of use, we unify these anomaly detectors following the same interface and wrap them up into a single Python package. Furthermore, we integrate our benchmark protocol into this package, which allows practitioners to automatically and comprehensively evaluate new anomaly detectors and compare them with existing state-of-the-art studies. 

Another target of our tool is reproducibility: the results could be reproduced by different teams with different setups~\cite{DBLP:conf/sigsoft/Hermann0S20}. To achieve this, in our open-source tool, we provide the dataset preprocessing scripts with corresponding processed datasets. Moreover, hyperparameters for different anomaly detectors are also released. Besides, we fix the random seed for the models producing indeterministic prediction (e.g., LSTM). 

\begin{table}[h]
    \caption{Dataset Statistics}
    \label{tab: dataset}
    \begin{tabular}{cccccc}
    \hline
    Dataset & \#Train & \#Test & \#Entities & \#Dim. & \begin{tabular}[c]{@{}c@{}}Anomaly \\ Ratio (\%)\end{tabular} \\ 
    \hline
    \hline
    SMD & 708,405 & 708,420 & 28 & 38 & 4.16 \\
    SMAP & 135,183 & 427,617 & 55 & 25 & 13.13 \\
    MSL & 58,317 & 73,729 & 27 & 55 & 10.72 \\
    SWAT & 496,800 & 449,919 & 1 & 40 & 12.14 \\
    WADI & 1,209,601 & 172,801 & 1 & 93 & 5.69 \\
    \hline
    \end{tabular}
    \end{table}

\section{Experiments}
\label{sec:experiments}

\begin{table*}[th]
    \centering
    \caption{Accuracy Comparison of Different Anomaly Detectors on Five real-world datasets: $F_1$ is calculated by using the threshold automatically selected via EVT, $\hat{F}_1$ denotes the F1 scores with the optimal threshold based on searching.}
    \includegraphics[scale=0.67]{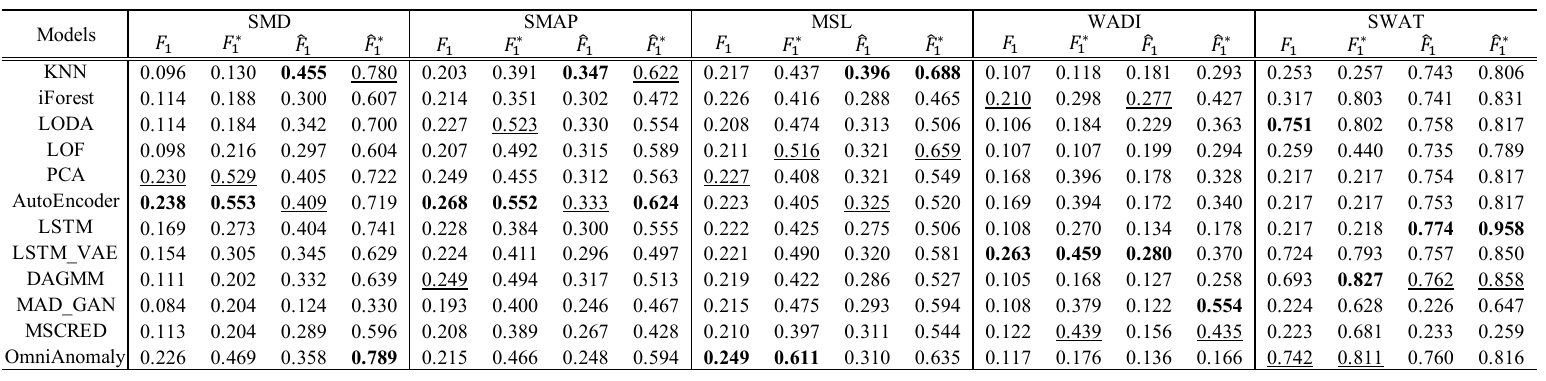}
    \label{fig: main_f1_table}
\end{table*}

In this section, we report our experimental results of state-of-the-art multivariate KPI anomaly detection models. Based on the proposed benchmark protocol, we aim to answer the following research questions (RQs).

\textbf{RQ1}: What is the accuracy of different methods?

\textbf{RQ2}: What is the salience of different methods?

\textbf{RQ3}: How timely can different methods detect anomalies?

\textbf{RQ4}: What is the efficiency of different methods?

\subsection{Experimental Setup}

\subsubsection{Dataset}
We adopt the following five widely used public datasets for our benchmark: 

SMD (Server Machine Dataset)~\cite{DBLP:conf/kdd/SuZNLSP19/omni}, which contains KPIs associated with 28 machines from a large Internet company.

SMAP (Soil Moisture Active Passive satellite) and MSL (Mars Science Laboratory rover~), which record the KPIs used to monitor the run-time information of a running spacecraft, which involves various spacecraft issues~\cite{DBLP:conf/kdd/HundmanCLCS18/lstm_ndt}, 

SWAT (Secure Water Treatment)~\cite{DBLP:conf/cpsweek/MathurT16} is collected from a real-world industrial water treatment plant~\cite{DBLP:conf/critis/GohAJM16}, which contains 11-day-long multivariate KPIs. Particularly, the system is in a normal state in the first seven days and is under attack in the following four days.

WADI (Water Distribution)~\cite{DBLP:conf/cpsweek/MathurT16} is an extended dataset of SWAT. 14-day-long operation KPIs are collected when the system is running normally and 2-day-long KPIs are obtained when the system is in attack scenarios.

In particular, Table~\ref{tab: dataset} shows the statistics of these datasets. The anomaly labels of these datasets are provided at the entity level. Hence, anomalies should be detected by considering all the multivariate KPIs as a whole. In addition, for SWAT and WADI datasets, we remove the KPIs that contain the same values, hence the ``\#Dim'' (i.e., number of dimension) of them are fewer than the original dataset.

\subsubsection{Experimental Environment}
We run all experiments on a Linux server with GeForce RTX 2080 Ti, Intel Xeon Gold 6148 CPU @ 2.40GHZ and 1TB RAM, running Red Hat 4.8.5 with Linux kernel 3.10.0. In addition, the deep learning-based models are executed with GPU acceleration.

\subsection{Accuracy (RQ1)}
Recent researches report the multivariate anomaly detection methods either reuse the results from previous studies or reproduce the baseline models~\cite{deng2021graph}~\cite{DBLP:conf/kdd/AudibertMGMZ20/usad}. However, the comparison may be unfair because of different hyper-parameters selection and evaluation protocol. In particular, the accuracy would be affected by using different threshold selection strategies and point adjustments.

In our benchmark protocol, we apply the evaluated anomaly detectors on all five datasets. Particularly, we adopt grid search to find the optimal hyper-parameter combination for each algorithm and report the best accuracy. We report the accuracy in terms of a different combination of threshold selection strategies and whether to use point adjustment. That is $F_1$ (by EVT threshold selection), $F_1^*$ (by EVT threshold selection and point adjustment), $\hat{F}_1$ (by threshold searching), $\hat{F}_1^*$ (by threshold searching and point adjustment). In particular, different from existing work~\cite{DBLP:conf/kdd/AudibertMGMZ20/usad}, while searching the optimal threshold, we use the accuracy $\hat{F}_1$ without point adjustment to supervise the procedure. In doing this, we prevent finding a threshold that leads to misleading higher accuracy mentioned in section~\ref{sec: motivation}.
Then, we apply the same threshold to obtain the adjusted version of accuracy, namely, $\hat{F}_1^*$. 

Table~\ref{fig: main_f1_table} shows the accuracy comparison of all the anomaly detection methods over the five datasets. For each metric (i.e., column), the highest value is marked as boldface, the second-highest one is underlined. We can make the following surprising observations: 
(1) Among all the datasets, $F_1$ is generally worse than $\hat{F}_1$, which indicates that EVT generally cannot find the optimal threshold (i.e., threshold obtained by search). Therefore, when anomaly labels are available, it is more practical to search the optimal threshold to best measure the accuracy of a model. Hence, the next observations are based on the accuracy obtained via search, i.e., $\hat{F}_1$ and $\hat{F}^*_1$.
(2) Deep learning-based models are not always the best solution. Surprisingly, KNN achieves the best performance in terms of $\hat{F}_1$ on SMD, SMAP and MSL. (3) After applying point adjustment, deep learning-based models present higher accuracies ($\hat{F}_1^*$). Specifically, except on MSL, the best $\hat{F}_1^*$ is achieved by deep learning-based methods, which implies that these methods are better at handling continuous anomalies.

\begin{figure}[h]
    \centering
    \includegraphics[scale=0.36]{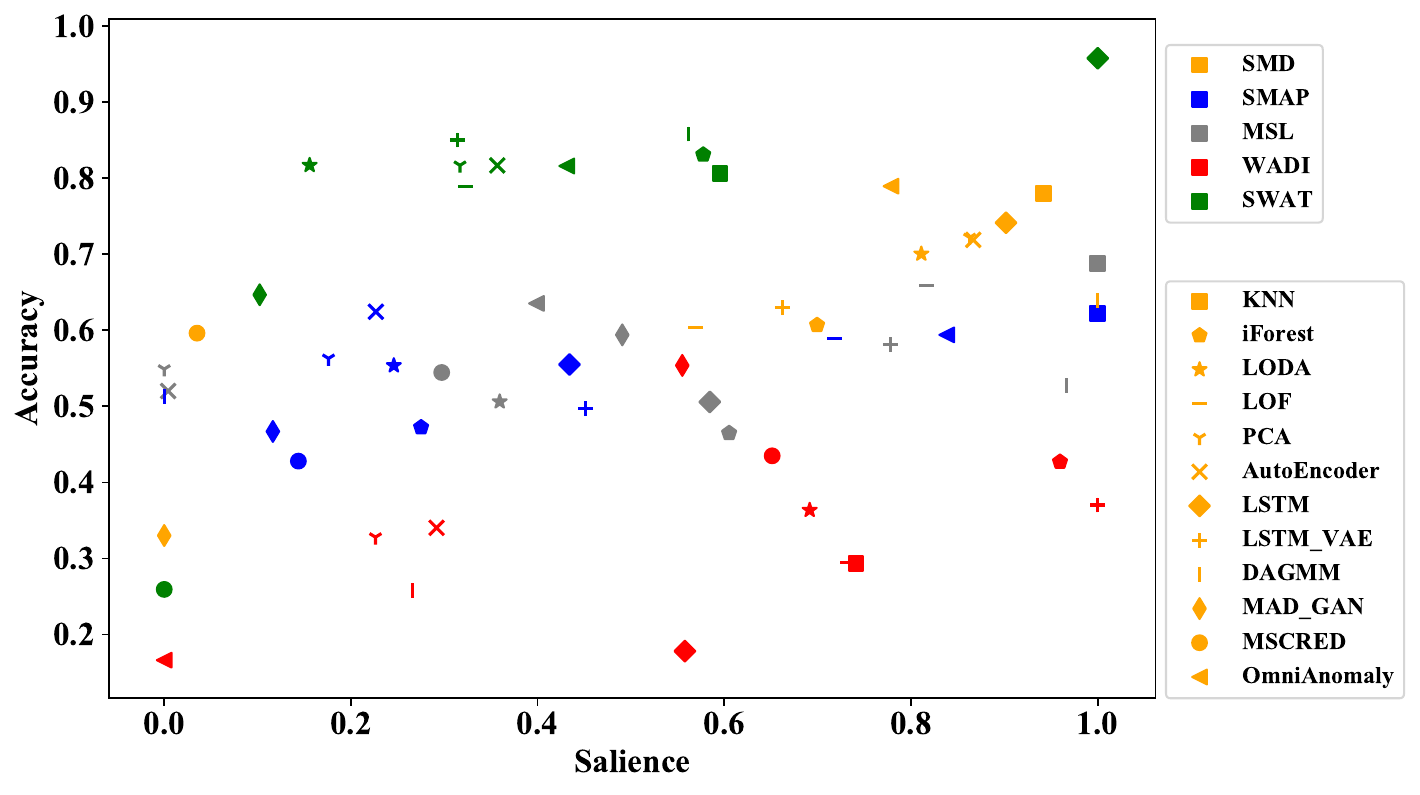}
    \caption{Model Performance in Terms of Accuracy and Salience} 
    \label{fig: salience_f1_relation}
    \end{figure}

\begin{figure*}[th]
    \centering
    \includegraphics[scale=0.38]{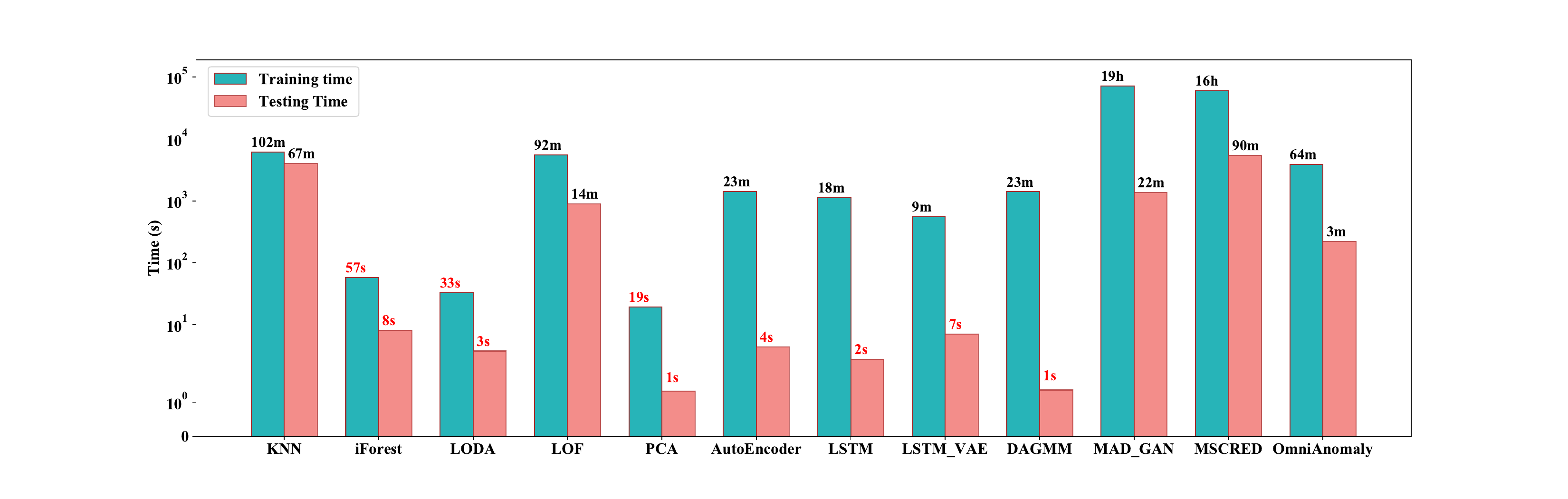}
    \caption{Training Time and Testing Time of Different Methods over WADI Dataset: Specific time cost is labled on the top of each bar, where those less than 1 minute are highlighted as red. We run general machine learning-based methods on CPU and deep learning-based methods on GPU.}
    \label{fig: efficiency}
    \end{figure*}

\subsection{Salience (RQ2)}
To further distinguish the more superior anomaly detector, we propose to use an additional novel metric salience as a supplement. In particular, salience measures how much an anomalous observation could be highlighted by an algorithm compared with normal observations.
We compute salience for 12 methods on 5 datasets. Within these datasets, SMD, SMAP, and MSL consist of more than one entity, hence, we compute salience for each of the entities and use the salience summation of all entities as the salience of the dataset. After that, we apply min-max normalization to the salience among the methods of each dataset.

Fig.~\ref{fig: salience_f1_relation} shows the relation between the accuracy and salience of our benchmark. In this figure, methods are denoted as the different shapes of markers, and different colors are used to distinguish different datasets. In particular, we use $\hat{F}_1^*$ as the accuracy axis, which indicates the best performance of a model. Moreover, the models near to the upper right corner are those with both high salience and accuracy.
We can find that models with high accuracy may not have high salience, which indicates that a threshold should be carefully selected for these models. Particularly, LSTM achieves the best accuracy and salience on SWAT. KNN also performs well in both metrics on SMD, SMAP, and MSL datasets. OmniAnomaly performs the worst on both metrics on the WADI dataset. To summarize, there is no model that could consistently perform well on all the datasets in terms of accuracy and salience.

\subsection{Delay (RQ3)}
Delay is another significant metric besides accuracy, which measures the time-lagged after an anomaly occurs. Delay is defined as the distance (i.e., number of points) between the first anomalous observation reported and the first true anomaly. An algorithm with lower delay would alert an anomaly timely, which can facilitate troubleshooting and alleviate potential system failure.
We compute delay based on the prediction vector $\hat{y}$ generated by using the optimal threshold $\theta_s$ obtained via search.
Table~\ref{tab: delay} shows the delay measurement of different methods, where the lowest delay is marked as boldface and the second-lowest one is underlined. In this table, we measure delay in terms of the number of points, hence, delay of WADI and SWAT may be larger than other datasets because they contain more observations. We can find that the lowest delay on all the datasets is generally achieved by traditional machine learning-based methods (i.e., KNN, LODA, lOF, and PCA). AutoEncoder and LSTM have 0 delays on WADI and SWAT, which is the same as PCA.  
The possible reason behind this is that traditional machine learning-method treat each observation as an independent point without considering historical dependency. As a result, these methods are more sensitive to the anomalies consisting of sudden spikes; hence, an outlier point could be immediately reported. Deep learning-based models like LSTM tend to take a long range of observations as the input. Hence, the prediction results might be smoothed such that anomalies may not be reported in a timely manner.
\begin{table}[th]
    \centering
    \caption{Delay Comparison of different models}
    \includegraphics[scale=0.68]{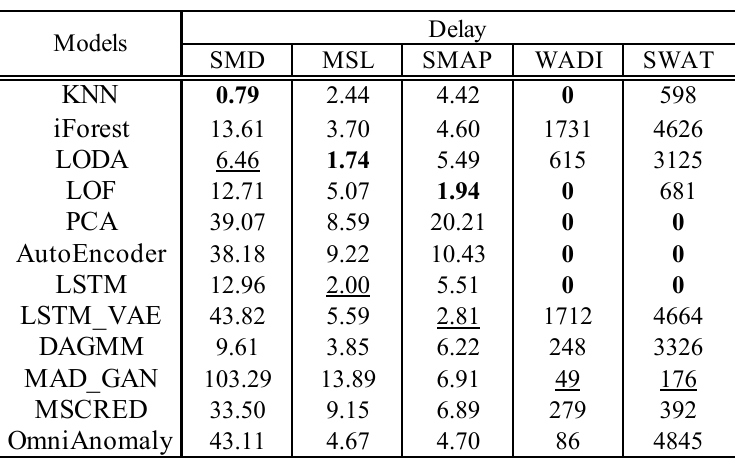}
    \label{tab: delay}
\end{table}

\subsection{Efficiency (RQ4)}
The efficiency of general machine learning-based or deep learning-based methods is evaluated in terms of training time and testing time. Specifically, training time indicates the time cost to train a model to convergence from scratch. High training efficiency allows a model to be updated frequently to capture new normal/anomaly patterns when systems upgrade. In addition, a model with high testing efficiency can timely process streaming input KPIs. Therefore, for an industry-level application, both training and testing efficiency is highly demanded.

To evaluate the efficiency of the anomaly detectors, we perform these methods on the WADI dataset, which contains the longest multivariate KPIs in total. To show the best efficiency of each model, we run the general machine learning-based methods on CPU and deep learning-based methods on GPU for acceleration. Particularly, the training time is calculated from the beginning of training to the model convergence. 
When the training loss of a model barely changes, we consider the model to be convergent. In addition, the testing time denotes the time cost to compute all the anomaly scores for the whole dataset. In particular, data loading/preprocessing time is not included for both training and testing time.

Fig.~\ref{fig: efficiency} shows the efficiency evaluation results for the anomaly detectors, where we mark the concrete time on the top of each bar for the models and highlight the timeless than one minute as red. We summarize the observations as:
(1) For general machine-learning methods, iForest, LODA, and PCA are the most efficient. KNN and LOF cost a lot more for both training and testing time. The reason is that KNN and LOF are based on the K-nearest neighbor search, where every point needs to make a comparison with all other points. Hence, when the volume of KPIs increases, the time cost will grow quadratically.
(2) For deep learning-based models, AutoEncoder, LSTM, LSTM\_VAE, DAGMM can make an efficient prediction and present similar training efficiency. Similarly, OmniAnomaly can be trained within nearly one hour and finish testing within 3 minutes. MAD\_GAN and MSCRED are the most inefficient methods, which require more than 15 hours to train. This could be explained by the design of the models: MAD\_GAN is based on a generative adversarial network (GAN), which needs a lot of computation. MSCRED needs to generate signature matrices which include pair-wise computation for all multivariate KPIs.  
(3) Deep learning-based models generally need more time to train with backpropagation, while can retain efficient testing, which benefits from the GPU acceleration.


\section{Related Work}
\label{sec: related_work}

Anomaly detection on time series has been a hot topic and is widely studied. 
Key performance indicators (KPIs) used to monitor the runtime status of a system are one of the classic types of time series. 
In the literature, there are two categories of scenarios. One scenario needs to conduct anomaly detection for individual KPI, which can be solved by univariate time series models efficiently~\cite{DBLP:conf/kdd/RenXWYHKXYTZ19/microsoft}.
Another scenario needs to consider the dependency among multiple KPIs to detect the anomalous status of the unified unit.


\textbf{Implicitly Modeling.} The major challenge of multivariate anomaly detection comes from the effective modeling of complex temporal dependence and stochasticity of Multivariate KPIs. 
Hundman et al.~\cite{DBLP:conf/kdd/HundmanCLCS18/lstm_ndt} leveraged LSTM without expert-labeled telemetry anomaly data to detect anomalies in multivariate time-series metrics of spacecraft based on prediction errors.
Malhotra et al. \cite{DBLP:journals/corr/MalhotraRAVAS16} proposed an LSTM-based encoder-decoder network to reconstruct the “normal” time series with high probabilities.
Another way to model normal patterns is to learn the distribution of input data like deep generative models~\cite{DBLP:conf/nips/GoodfellowPMXWOCB14} and deep Bayesian network~\cite{DBLP:journals/tkde/WangY16}. 
Donut~\cite{DBLP:conf/www/XuCZLBLLZPFCWQ18/donut} employed Variational AutoEncoder (VAE) to generate the normal hidden state of seasonal KPIs without expert-labeled data.
Donut successfully detects anomalies in seasonal KPIs with various patterns and data quality, but enjoys high time complexity in the training phase.
To reduce the training complexity, DAGMM~\cite{DBLP:conf/iclr/ZongSMCLCC18/dagmm} simplifies the hidden state as a combination of several Gaussian distributions.
USAD~\cite{DBLP:conf/kdd/AudibertMGMZ20/usad} improves the autoencoder framework by incorporating adversarial samples to speed up the training phase.
OmniAnomaly~\cite{DBLP:conf/kdd/SuZNLSP19/omni} employs a stochastic recurrent neural network to captures the normal patterns of multivariate time-series by simulating normal data distribution through stochastic latent variables.
Similar to Hundman et al.~\cite{DBLP:conf/kdd/HundmanCLCS18/lstm_ndt}'s approach, OmniAnomaly provides interpretations based on the reconstruction probabilities of its constituent univariate KPI.
However, the generalization capability of these generative approaches with implicitly modeling is degraded when they encounter severe noise in temporal KPIs, which is very common in industrial production systems. 

\textbf{Explicitly Modeling.} To reduce the negative effect of noisy data in multivariate KPIs, the temporal dependencies such as the inter-correlations between different parts of KPIs should be captured in an explicit way. 
Zhang et al.~\cite{DBLP:conf/aaai/ZhangSCFLCNZCC19} proposed a convolutional recurrent encoder-decoder framework called MSCRED to characterize multiple levels of the normal states in different steps of KPIs.
MSCRED utilizes a convolutional encoder and convolutional LSTM network to capture the KPI interactions and temporal patterns, respectively.
Zhao et al.~\cite{DBLP:conf/icdm/ZhaoWDHCTXBTZ20} considered each univariate KPI as an individual feature to capture the complex dependencies of multivariate KPIs from temporal and feature perspective.
Their proposed method incorporates feature-oriented and time-oriented graph attention mechanisms to obtain mixed hidden representation from a combination of forecasting-based and reconstruction-based frameworks.
However, the high time complexity in the overall framework and the carefully selected thresholds about the anomaly persist barriers of demanding industrial requirements for quick response and adjustment.
\section{Conclusion}
\label{sec:conclusion}
In this work, we argue that current multivariate KPIs anomaly detection studies are evaluated separately instead of a uniform protocol. Moreover, existing work tends to estimate only the accuracy of a model. As a consequence, it is non-trivial to reproduce the results in the literature and select a suitable method for industrial deployment. To address this problem, we propose a comprehensive benchmark protocol with a novel metric \textit{salience}. Under this uniform protocol, we re-evaluate twelve state-of-the-art multivariate KPIs anomaly detection on five real-world public datasets and report the experimental results in terms of accuracy, salience, efficiency, and delay. We also released an easy-to-use toolkit that automatically evaluates multivariate KPIs anomaly detection with our protocol. Hence, we believe this can benefit practitioners from both academia and industry to develop advanced anomaly detectors.

\bibliographystyle{IEEEtran}
\bibliography{ase21}

\begin{thebibliography}{10}
\providecommand{\url}[1]{#1}
\csname url@samestyle\endcsname
\providecommand{\newblock}{\relax}
\providecommand{\bibinfo}[2]{#2}
\providecommand{\BIBentrySTDinterwordspacing}{\spaceskip=0pt\relax}
\providecommand{\BIBentryALTinterwordstretchfactor}{4}
\providecommand{\BIBentryALTinterwordspacing}{\spaceskip=\fontdimen2\font plus
\BIBentryALTinterwordstretchfactor\fontdimen3\font minus \fontdimen4\font\relax}
\providecommand{\BIBforeignlanguage}[2]{{%
\expandafter\ifx\csname l@#1\endcsname\relax
\typeout{** WARNING: IEEEtran.bst: No hyphenation pattern has been}%
\typeout{** loaded for the language `#1'. Using the pattern for}%
\typeout{** the default language instead.}%
\else
\language=\csname l@#1\endcsname
\fi
#2}}
\providecommand{\BIBdecl}{\relax}
\BIBdecl

\bibitem{DBLP:conf/sigsoft/ChenKLZZXZYSXDG20}
Z.~Chen, Y.~Kang, L.~Li, X.~Zhang, H.~Zhang, H.~Xu, Y.~Zhou, L.~Yang, J.~Sun, Z.~Xu, Y.~Dang, F.~Gao, P.~Zhao, B.~Qiao, Q.~Lin, D.~Zhang, and M.~R. Lyu, ``Towards intelligent incident management: why we need it and how we make it,'' in \emph{Proceedings of the 28th Joint European Software Engineering Conference and Symposium on the Foundations of Software Engineering, (ESEC/FSE)}, 2020, pp. 1487--1497.

\bibitem{DBLP:conf/usenix/ZhangLXQ0QDYCCW19}
X.~Zhang, Q.~Lin, Y.~Xu, S.~Qin, H.~Zhang, B.~Qiao, Y.~Dang, X.~Yang, Q.~Cheng, M.~Chintalapati, Y.~Wu, K.~Hsieh, K.~Sui, X.~Meng, Y.~Xu, W.~Zhang, F.~Shen, and D.~Zhang, ``Cross-dataset time series anomaly detection for cloud systems,'' in \emph{Proceedings of the Annual Technical Conference, (USENIX ATC)}, 2019, pp. 1063--1076.

\bibitem{DBLP:conf/icse/DangLH19}
Y.~Dang, Q.~Lin, and P.~Huang, ``Aiops: real-world challenges and research innovations,'' in \emph{Proceedings of the 41st International Conference on Software Engineering: Companion Proceedings, (ICSE)}, 2019, pp. 4--5.

\bibitem{DBLP:journals/corr/abs-2009-07237}
S.~He, P.~He, Z.~Chen, T.~Yang, Y.~Su, and M.~R. Lyu, ``A survey on automated log analysis for reliability engineering,'' \emph{CoRR}, vol. abs/2009.07237, 2020.

\bibitem{DBLP:conf/sigsoft/HeLLZLZ18}
S.~He, Q.~Lin, J.~Lou, H.~Zhang, M.~R. Lyu, and D.~Zhang, ``Identifying impactful service system problems via log analysis,'' in \emph{Proceedings of the Joint Meeting on European Software Engineering Conference and Symposium on the Foundations of Software Engineering, (ESEC/FSE) {FSE}}.\hskip 1em plus 0.5em minus 0.4em\relax {ACM}, 2018, pp. 60--70.

\bibitem{gu2023performance}
W.~Gu, J.~Liu, Z.~Chen, J.~Zhang, Y.~Su, J.~Gu, C.~Feng, Z.~Yang, and M.~Lyu, ``Performance issue identification in cloud systems with relational-temporal anomaly detection,'' \emph{arXiv preprint arXiv:2307.10869}, 2023.

\bibitem{DBLP:conf/kdd/HundmanCLCS18/lstm_ndt}
K.~Hundman, V.~Constantinou, C.~Laporte, I.~Colwell, and T.~S{\"{o}}derstr{\"{o}}m, ``Detecting spacecraft anomalies using lstms and nonparametric dynamic thresholding,'' in \emph{Proceedings of the 24th International Conference on Knowledge Discovery {\&} Data Mining, (KDD)}, 2018, pp. 387--395.

\bibitem{borghesi2019anomaly}
A.~Borghesi, A.~Bartolini, M.~Lombardi, M.~Milano, and L.~Benini, ``Anomaly detection using autoencoders in high performance computing systems,'' in \emph{Proceedings of the AAAI Conference on Artificial Intelligence}, vol.~33, no.~01, 2019, pp. 9428--9433.

\bibitem{DBLP:journals/corr/abs-1711-00614/lstm-vae}
D.~Park, Y.~Hoshi, and C.~C. Kemp, ``A multimodal anomaly detector for robot-assisted feeding using an lstm-based variational autoencoder,'' \emph{CoRR}, vol. abs/1711.00614, 2017.

\bibitem{li2019mad}
D.~Li, D.~Chen, B.~Jin, L.~Shi, J.~Goh, and S.-K. Ng, ``Mad-gan: Multivariate anomaly detection for time series data with generative adversarial networks,'' in \emph{International Conference on Artificial Neural Networks}.\hskip 1em plus 0.5em minus 0.4em\relax Springer, 2019, pp. 703--716.

\bibitem{DBLP:conf/aaai/ZhangSCFLCNZCC19}
C.~Zhang, D.~Song, Y.~Chen, X.~Feng, C.~Lumezanu, W.~Cheng, J.~Ni, B.~Zong, H.~Chen, and N.~V. Chawla, ``A deep neural network for unsupervised anomaly detection and diagnosis in multivariate time series data,'' in \emph{Proceedings of the 33rd Applications of Artificial Intelligence Conference, (AAAI)}, 2019, pp. 1409--1416.

\bibitem{siffer2017anomaly}
A.~Siffer, P.-A. Fouque, A.~Termier, and C.~Largouet, ``Anomaly detection in streams with extreme value theory,'' in \emph{Proceedings of the 23rd ACM SIGKDD International Conference on Knowledge Discovery and Data Mining}, 2017, pp. 1067--1075.

\bibitem{DBLP:conf/www/XuCZLBLLZPFCWQ18/donut}
H.~Xu, W.~Chen, N.~Zhao, Z.~Li, J.~Bu, Z.~Li, Y.~Liu, Y.~Zhao, D.~Pei, Y.~Feng, J.~Chen, Z.~Wang, and H.~Qiao, ``Unsupervised anomaly detection via variational auto-encoder for seasonal kpis in web applications,'' in \emph{Proceedings of the 2018 World Wide Web Conference on World Wide Web, (WWW)}.\hskip 1em plus 0.5em minus 0.4em\relax {ACM}, 2018, pp. 187--196.

\bibitem{DBLP:conf/icdm/ZhaoWDHCTXBTZ20}
H.~Zhao, Y.~Wang, J.~Duan, C.~Huang, D.~Cao, Y.~Tong, B.~Xu, J.~Bai, J.~Tong, and Q.~Zhang, ``Multivariate time-series anomaly detection via graph attention network,'' in \emph{20th {IEEE} International Conference on Data Mining, {ICDM} 2020, Sorrento, Italy, November 17-20, 2020}.\hskip 1em plus 0.5em minus 0.4em\relax {IEEE}, 2020, pp. 841--850.

\bibitem{deng2021graph}
A.~Deng and B.~Hooi, ``Graph neural network-based anomaly detection in multivariate time series,'' in \emph{Proceedings of the 35th AAAI Conference on Artificial Intelligence}, 2021.

\bibitem{DBLP:journals/corr/abs-2004-00433/univariate_survey}
M.~Braei and S.~Wagner, ``Anomaly detection in univariate time-series: {A} survey on the state-of-the-art,'' \emph{CoRR}, vol. abs/2004.00433, 2020.

\bibitem{ramaswamy2000efficient}
S.~Ramaswamy, R.~Rastogi, and K.~Shim, ``Efficient algorithms for mining outliers from large data sets,'' in \emph{Proceedings of the 2000 ACM SIGMOD international conference on Management of data}, 2000, pp. 427--438.

\bibitem{breunig2000lof}
M.~M. Breunig, H.-P. Kriegel, R.~T. Ng, and J.~Sander, ``Lof: identifying density-based local outliers,'' in \emph{Proceedings of the 2000 ACM SIGMOD international conference on Management of data}, 2000, pp. 93--104.

\bibitem{DBLP:conf/icdm/LiuTZ08/isolation_forest}
F.~T. Liu, K.~M. Ting, and Z.~Zhou, ``Isolation forest,'' in \emph{Proceedings of the 8th International Conference on Data Mining (ICDM)}, 2008, pp. 413--422.

\bibitem{pevny2016loda}
T.~Pevn{\`y}, ``Loda: Lightweight on-line detector of anomalies,'' \emph{Machine Learning}, vol. 102, no.~2, pp. 275--304, 2016.

\bibitem{DBLP:conf/kdd/SuZNLSP19/omni}
Y.~Su, Y.~Zhao, C.~Niu, R.~Liu, W.~Sun, and D.~Pei, ``Robust anomaly detection for multivariate time series through stochastic recurrent neural network,'' in \emph{Proceedings of the 25th International Conference on Knowledge Discovery {\&} Data Mining, (KDD)}, 2019, pp. 2828--2837.

\bibitem{DBLP:conf/iclr/ZongSMCLCC18/dagmm}
B.~Zong, Q.~Song, M.~R. Min, W.~Cheng, C.~Lumezanu, D.~Cho, and H.~Chen, ``Deep autoencoding gaussian mixture model for unsupervised anomaly detection,'' in \emph{Proceedings of the 6th International Conference on Learning Representations, (ICLR)}, 2018.

\bibitem{ren2019time}
H.~Ren, B.~Xu, Y.~Wang, C.~Yi, C.~Huang, X.~Kou, T.~Xing, M.~Yang, J.~Tong, and Q.~Zhang, ``Time-series anomaly detection service at microsoft,'' in \emph{Proceedings of the 25th ACM SIGKDD International Conference on Knowledge Discovery \& Data Mining}, 2019, pp. 3009--3017.

\bibitem{DBLP:conf/kdd/AudibertMGMZ20/usad}
J.~Audibert, P.~Michiardi, F.~Guyard, S.~Marti, and M.~A. Zuluaga, ``{USAD:} unsupervised anomaly detection on multivariate time series,'' in \emph{Proceedings of the 26th {SIGKDD} Conference on Knowledge Discovery and Data Mining, (KDD)}.\hskip 1em plus 0.5em minus 0.4em\relax {ACM}, 2020, pp. 3395--3404.

\bibitem{DBLP:conf/kdd/RenXWYHKXYTZ19/microsoft}
H.~Ren, B.~Xu, Y.~Wang, C.~Yi, C.~Huang, X.~Kou, T.~Xing, M.~Yang, J.~Tong, and Q.~Zhang, ``Time-series anomaly detection service at microsoft,'' in \emph{Proceedings of the 25th International Conference on Knowledge Discovery {\&} Data Mining, (KDD)}, 2019, pp. 3009--3017.

\bibitem{DBLP:conf/issre/HeZHL16/loglizer}
S.~He, J.~Zhu, P.~He, and M.~R. Lyu, ``Experience report: System log analysis for anomaly detection,'' in \emph{Proceedings of the 27th International Symposium on Software Reliability Engineering, (ISSRE)}, 2016, pp. 207--218.

\bibitem{DBLP:conf/ccs/Du0ZS17/deeplog}
M.~Du, F.~Li, G.~Zheng, and V.~Srikumar, ``Deeplog: Anomaly detection and diagnosis from system logs through deep learning,'' in \emph{Proceedings of the 2017 Conference on Computer and Communications Security, (CCS)}.\hskip 1em plus 0.5em minus 0.4em\relax {ACM}, 2017, pp. 1285--1298.

\bibitem{DBLP:conf/kdd/SifferFTL17/EVT}
A.~Siffer, P.~Fouque, A.~Termier, and C.~Largou{\"{e}}t, ``Anomaly detection in streams with extreme value theory,'' in \emph{Proceedings of the 23rd {SIGKDD} International Conference on Knowledge Discovery and Data Mining, (KDD)}.\hskip 1em plus 0.5em minus 0.4em\relax {ACM}, 2017, pp. 1067--1075.

\bibitem{wiki_complete_linkage}
Wikipedia, ``{Complete-linkage},'' \url{http://en.wikipedia.org/wiki/Complete-linkage\_clustering}, 2021, [Online; accessed 23-April-2021].

\bibitem{zhao2019pyod}
\BIBentryALTinterwordspacing
Y.~Zhao, Z.~Nasrullah, and Z.~Li, ``Pyod: A python toolbox for scalable outlier detection,'' \emph{Journal of Machine Learning Research}, vol.~20, no.~96, pp. 1--7, 2019. [Online]. Available: \url{http://jmlr.org/papers/v20/19-011.html}
\BIBentrySTDinterwordspacing

\bibitem{DBLP:journals/tkde/LuLDGGZ19}
J.~Lu, A.~Liu, F.~Dong, F.~Gu, J.~Gama, and G.~Zhang, ``Learning under concept drift: {A} review,'' \emph{{IEEE} Trans. Knowl. Data Eng.}, vol.~31, no.~12, pp. 2346--2363, 2019.

\bibitem{DBLP:conf/sigsoft/Hermann0S20}
B.~Hermann, S.~Winter, and J.~Siegmund, ``Community expectations for research artifacts and evaluation processes,'' in \emph{{ESEC/FSE} '20: 28th {ACM} Joint European Software Engineering Conference and Symposium on the Foundations of Software Engineering, Virtual Event, USA, November 8-13, 2020}, P.~Devanbu, M.~B. Cohen, and T.~Zimmermann, Eds.\hskip 1em plus 0.5em minus 0.4em\relax {ACM}, 2020, pp. 469--480.

\bibitem{DBLP:conf/cpsweek/MathurT16}
A.~P. Mathur and N.~O. Tippenhauer, ``Swat: a water treatment testbed for research and training on {ICS} security,'' in \emph{2016 International Workshop on Cyber-physical Systems for Smart Water Networks, CySWater@CPSWeek 2016, Vienna, Austria, April 11, 2016}.\hskip 1em plus 0.5em minus 0.4em\relax {IEEE} Computer Society, 2016, pp. 31--36.

\bibitem{DBLP:conf/critis/GohAJM16}
J.~Goh, S.~Adepu, K.~N. Junejo, and A.~Mathur, ``A dataset to support research in the design of secure water treatment systems,'' in \emph{Critical Information Infrastructures Security - 11th International Conference, {CRITIS} 2016, Paris, France, October 10-12, 2016, Revised Selected Papers}, ser. Lecture Notes in Computer Science.\hskip 1em plus 0.5em minus 0.4em\relax Springer, 2016, pp. 88--99.

\bibitem{DBLP:journals/corr/MalhotraRAVAS16}
P.~Malhotra, A.~Ramakrishnan, G.~Anand, L.~Vig, P.~Agarwal, and G.~Shroff, ``Lstm-based encoder-decoder for multi-sensor anomaly detection,'' \emph{CoRR}, vol. abs/1607.00148, 2016.

\bibitem{DBLP:conf/nips/GoodfellowPMXWOCB14}
I.~J. Goodfellow, J.~Pouget{-}Abadie, M.~Mirza, B.~Xu, D.~Warde{-}Farley, S.~Ozair, A.~C. Courville, and Y.~Bengio, ``Generative adversarial nets,'' in \emph{Proceedings of the 27th Conference on Neural Information Processing Systems 2014, (NeurIPS)}, 2014, pp. 2672--2680.

\bibitem{DBLP:journals/tkde/WangY16}
H.~Wang and D.~Yeung, ``Towards bayesian deep learning: {A} framework and some existing methods,'' \emph{{IEEE} Trans. Knowl. Data Eng.}, vol.~28, no.~12, pp. 3395--3408, 2016.

\end{thebibliography}
\end{document}